Research Article

# Can renewable energy sources alleviate the pressure of military expenditures on the environment? Empirical evidence from Turkiye

Emre AKUSTA[1]

[1]*Kirklareli University, Department of Economics, Kofcaz, 39100 Kirklareli, Türkiye*

**ABSTRACT**

This study analyzes the potential of renewable energy sources to reduce the environmental impact of military expenditures in Turkiye. ARDL method is preferred in the analysis using annual data for the period 1990-2021. In addition, an interaction term is added to the model to determine the effectiveness of renewable energy sources. The results show that military expenditures have a positive impact on $CO_2$ emissions in the short and long run with coefficients of 0.260 and 0.196, respectively. Moreover, renewable energy use has a statistically significant negative impact on $CO_2$ emissions in the short and long run with coefficients of -0.119 and -0.120, respectively. GDP has a positive impact on $CO_2$ emissions in the short and long run with coefficients of 0.162 and 0.193, respectively. Although population growth does not have a statistically significant impact in the short run, it is found to increase $CO_2$ emissions in the long run with a coefficient of 0.095. Moreover, the interaction term shows that renewable energy use reduces the environmental impact of military expenditures in Turkiye in the short and long run with coefficients of -0.130 and -0.140, respectively. The results indicate that renewable energy use can play an important role in mitigating the environmental impacts of military expenditures.

**Cite this article as:** Akusta E. Can renewable energy sources alleviate the pressure of military expenditures on the environment? Empirical evidence from Turkiye. Environ Res Tec 2025;8(2) 410-424.

## INTRODUCTION

Security and defense have been one of the most important priorities of states throughout human history. This need is important for societies to sustain their existence, become a regional power and access new sources of wealth. This is described by [1] as states constantly strengthening their defense mechanisms in order to protect their economic welfare and ensure their security. However, in the modern world, this process is no longer limited to military capabilities. Economic dynamics, security challenges and environmental impacts determine defense strategies and policies.

The widespread use of oil after the industrial revolution increased the demand for energy and accelerated the industrialization and economic growth processes of countries. In this period, the results of wars were determined not by the number of soldiers but by technical weapons and logistical support. Therefore, having strong economies and industrialization became critical for defense [1], [2]. Stable economic growth requires industrialization and defense activities. Economic progress cannot be sustained without a developed industrial infrastructure and a strong defense mechanism. These three factors are closely integrated. Industrialization supports economic growth through its capacity to both create jobs and provide greater economic diversification. At the same time, advanced industrial production provides the infrastructure and technology needed to develop defense technologies. In addition to ensuring national security, a strong defense sector contributes to economic growth by stimulating high-tech research and innovation. This synergy is more apparent in developed countries. In underdeveloped countries, however, excessive military expenditures can have a negative impact on economic growth [3], [4]. However, these activities are not limited to military capacity, but are directly linked to the continuity of energy resources. Industrial activities, defense operations and economic growth cannot be sustained without energy resources. The importance of energy leads states to increase their defense expenditures in order to access or protect energy resources. This makes the interaction between energy and defense one of the most important economic and political issues for states. Therefore, there is a strong interaction between energy policies and defense expenditures. Military expenditure, which is intertwined with industrialization and economic growth, may also conflict with environmental sustainability goals. In this context, policies at the national and global level are central to the sustainability of economic activities, access to energy resources and defense strategies.

---

*Corresponding author.
*E-mail address: emre.akusta@klu.edu.tr

Interest in renewable energy sources has increased due to their potential to reduce environmental destruction and increase energy security. Therefore, global energy policies have been revised to reduce greenhouse gas emissions and increase the use of renewable energy sources. These steps have been supported by international agreements and protocols. For example, agreements such as the United Nations Framework Convention on Climate Change and the Kyoto Protocol have formalized countries' commitments to reduce greenhouse gas emissions. Renewable energy sources are also an important alternative for energy-dependent and resource-limited countries. By replacing limited fossil fuels, these resources can increase countries' energy security and improve their economic balance. Furthermore, the report by [5] estimates that renewable energy sources will account for two-thirds of the total energy supply by 2050. This plays a critical role in reducing energy-related $CO_2$ emissions and limiting global temperature rise.

Sustainable development can be defined as the process of meeting the needs of the present without compromising the ability of future generations to meet their needs. By supporting environmental sustainability in this process, renewable energy sources can provide many benefits such as increasing energy security and reducing environmental pollution. Moreover, energy R&D expenditures and innovative technologies can make significant contributions to increasing renewable energy production and reducing costs [6]. In this regard, integrating renewable energy policies with military strategies would be an important step in terms of both economic and environmental sustainability. However, the environmental impacts of military expenditures also pose a challenge for sustainable development. The environmental costs of military activities and the defense industry are one of the obstacles to sustainable growth. In this respect, a green growth approach can play a critical role in balancing both the economic and environmental impacts of military expenditures. Therefore, the impact of military expenditures on environmental quality should be reassessed and the use of renewable energy sources should be encouraged. This approach can both reduce environmental damage and support energy security and economic stability. To this end, this study investigates whether renewable energy sources can reduce the environmental impacts of military expenditures in Turkiye.

This study will contribute to the literature in at least three ways. Firstly, previous studies have generally investigated the impacts of military expenditures or renewable energy on the environment independently. However, there has been no analysis on the interactions between these two factors. To the best of our knowledge, this is the first study to investigate the potential of renewable energy sources to reduce the environmental impacts of military expenditures in Turkiye. Secondly, this study uses the interaction term methodology to analyze the interaction between military expenditures and renewable energy use in Turkiye. This methodology was chosen specifically to examine the interactive impacts of these two variables on the environment, an approach that is not often used in the Turkish context. Finally, this study is conducted using the most recent datasets available in Turkiye.

The rest of the paper is organized as follows: Section 2 presents military expenditures and $CO_2$ emissions in historical perspective, Section 3 presents the literature review, Section 4 presents the data and methodology, Section 5 presents the empirical findings and finally Section 6 presents the conclusions and policy implications.

## MILITARY EXPENDITURES AND $CO_2$ EMISSIONS IN HISTORICAL PERSPECTIVE

Global warming and environmental pollution are one of the greatest challenges of today's world. The use of fossil fuels and industrial activities are often at the root of environmental problems, which can have serious consequences such as climate change and the degradation of ecosystems. However, another important but often overlooked source of environmental pollution is military expenditures and activities. Military expenditures are often associated with high energy consumption and fossil fuel use. This can trigger environmental pollution both directly and indirectly by increasing $CO_2$ emissions.

Military activities include a wide range of energy-intensive operations. These operations range from logistical support to the use of fighter aircraft, which consume high amounts of fossil fuels. Moreover, the construction and maintenance of military infrastructures also requires significant consumption of energy and resources. Obviously, understanding the impact of military expenditures on $CO_2$ emissions and their contribution to environmental pollution is an important area of research in line with sustainable development goals. Moreover, while many countries increase their defense budgets, they may ignore the environmental impacts of these expenditures. Increases in defense budgets are often in direct conflict with environmental policies [7]. Therefore, assessing the environmental impacts of military expenditures and developing strategies to minimize these impacts are of great importance for both national and global environmental poli-cies. In this regard, the historical development of military expenditures in Turkiye and the world is presented in Figure 1.

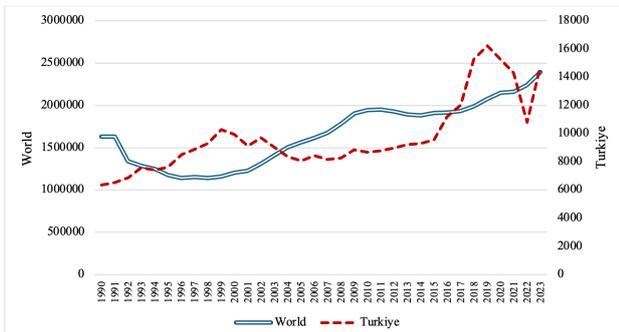

**Figure 1**. Evolution of military expenditures. Source: Constructed with data from [8].

Figure 1 shows military expenditures in Turkiye and the world in constant thousand dollars. Figure 1 shows that there was a general downward trend in military expenditures worldwide in the 1990-2000 period. Military expenditures decreased from approximately 1.6 trillion dollars in 1990 to approximately 1.2 trillion dollars in 2000. Factors such as the end of the Cold War and the dissolution of the Eastern Bloc were influential in the decline in expenditures during this period. These changes led many countries to cut their defense budgets and reduce their military expenditures. However, local conflicts and civil wars in some regions caused military expenditures to remain high on a local basis. Since 2001, military expenditures have started to increase again worldwide. While military expenditures were approximately 1.22 trillion dollars in 2001, they increased to approximately 1.94 trillion dollars in 2010. In this period, especially the counter-terrorism operations launched after the September 11 attacks and the conflicts in the Middle East were decisive in the increase in military expenditures. Many countries have sought to increase their defense capacities against security threats. The global war on terrorism led to significant increases in the defense budgets of the US and its allies [9], [10]. Military expenditures remained at high levels in the 2011-2023 period. Reaching 2.16 trillion dollars in 2021, expenditures increased to 2.39 trillion dollars in 2023. In this period, the increase in global security threats and the continuation of military conflicts in various regions caused expenditures to remain at high levels. During this period, China's efforts to increase its military power and events such as Russia's intervention in Ukraine were effective in increasing military expenditures [11]. At the same time, investments in new areas such as cyber security and space defense have also contributed to the increase in military budgets.

Figure 1 shows that Turkiye's military expenditures have generally fluctuated against the global trend. In the 1990-2000 period, Turkiye's military expenditures showed a general upward trend. From approximately 6.85 billion dollars in 1992, military expenditures reached 10.27 billion dollars in 1999. During this period, Turkiye's fight against terrorism was an important factor in the increase in military expenditures. Turkiye maintained a strong defense policy against internal security threats and accordingly increased its military expenditures [12], [13]. Turkiye's military expenditures fluctuated between 2001 and 2010. From 9.11 billion dollars in 2001, expenditures remained around 8.69 billion dollars in 2010. Economic crises and financial discipline were effective in the fluctuating course of expenditures in this period. In this period, Turkiye made adjustments in its defense budget in order to ensure economic stability. In the 2011-2023 period, a significant upward trend was observed in Turkiye's military expenditures. From USD 12.03 billion in 2017, expenditures reached USD 15.27 billion in 2018 and USD 14.32 billion in 2021. However, expenditures were recorded as USD 14.74 billion in 2023. Meanwhile, Turkiye's regional security policies, investments in the defense industry and operations in Syria played an important role in the increase in military expenditures. Turkiye took important steps to develop its own defense industry, which led to an increase in military expenditures. In particular, domestically produced weapons and equipment projects have played an important role in Turkiye's military budget. Moreover, Turkiye's efforts to protect its interests in the Eastern Mediterranean and disputes over energy resources in the region have also contributed to the increase in military expenditures.

After examining the historical evolution of military expenditures, it is important to review the historical evolution of $CO_2$ emissions in order to better understand the environmental impacts of these expenditures. Therefore, Figure 2 illustrates the historical development of $CO_2$ emissions for both Turkiye and the world.

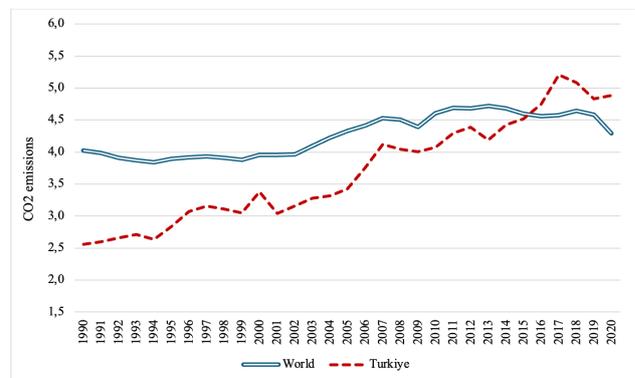

**Figure 2**. $CO_2$ emissions in Turkiye and the world. Source: Constructed with data from [14].

Figure 2 shows $CO_2$ emissions in Turkiye and the world in metric tons per capita. In the 1990-2000 period, $CO_2$ emissions per capita in the world generally followed a fluctuating trend. $CO_2$ emissions, which were about 4.02 metric tons in 1990, declined to about 3.95 metric tons in 2000. During this period, increased environmental awareness and straitening of environmental regulations in some industrialized countries were effective in reducing emissions. In particular, the signing of the Kyoto Protocol strengthened countries' commitments to reduce carbon emissions. However, increasing energy demand and intensive use of fossil fuels in developing countries limited the decline in global carbon emissions. The impact of global economic growth in this period should also be taken into account. Economic growth leads to an increase in energy consumption and thus higher carbon emissions.

However, the adoption of sustainable energy policies and investments in renewable energy sources can help offset this impact. Between 2001 and 2010, $CO_2$ emissions worldwide started to increase again. Emissions of about 3.96 metric tons in 2001 increased to about 4.60 metric tons in 2010. During this period, the increase in energy consumption and the widespread use of fossil fuels, especially in countries with rapid economic growth such as China and India, have been decisive in the increase in global $CO_2$ emissions [15]. In addition, high energy demand in developed countries such as the US has also contributed to the increase in global emissions. Some developed countries have continued their efforts to reduce carbon emissions through energy efficiency projects and renewable energy investments. However, the dominance of fossil fuels in energy production continued [16], [17]. On the other hand, $CO_2$ emissions worldwide followed a fluctuating pattern in the 2011-2020 period. Emissions, which were approximately 4.69 metric tons in 2011, decreased to approximately 4.29 metric tons in 2020. In this period, the increase in the use of renewable energy sources and energy efficiency measures were effective in reducing emissions. Moreover, with the signing of the Paris Agreement, many countries strengthened their commitments to reduce carbon emissions. However, the continued use of fossil fuels in some regions and the unabated industrialization caused global emissions to remain at high levels. The economic recession, especially during the COVID-19 pandemic, led to a decline in emissions in 2020, but this decline was not permanent [18].

In the 1990-2000 period, Turkiye's per capita $CO_2$ emissions showed a general upward trend. $CO_2$ emissions, which were about 2.56 metric tons in 1990, increased to about 3.37 metric tons in 2000. During this period, Turkiye's industrialization and urbanization processes accelerated and the increase in energy consumption and fossil fuel use played an important role in the increase in emissions. In particular, the widespread use of fossil fuels such as coal and natural gas in energy production contributed to the increase in Turkiye's carbon emissions. However, Turkiye's efforts to invest in renewable energy sources and implement energy efficiency projects can contribute to reducing carbon emissions in the long run [19]. Turkiye's $CO_2$ emissions followed a volatile trend in the period 2001-2010. Emissions increased from about 3.04 metric tons in 2001 to about 4.07 metric tons in 2010. In this period, Turkiye's economic growth and the increase in energy demand were the determinants of the increase in carbon emissions. The increased use of fossil fuels, especially in the industrial and transportation sectors, led to an increase in emissions. However, investments in renewable energy sources and energy efficiency projects have tried to limit the rate of increase in emissions. The widespread use of fossil fuels in the industrial and transportation sectors in Turkiye has been an important factor in the increase in carbon emissions [20]. During this period, shifting Turkiye's energy policies towards renewable resources and adopting environmental sustainability policies, expanding energy efficiency projects and reducing fossil fuel use can help reduce carbon emissions. In the period 2011-2020, Turkiye's $CO_2$ emissions showed a general upward trend. Emissions increased from about 4.29 metric tons in 2011 to about 4.88 metric tons in 2020. Throughout this period, Turkiye's efforts to transition to renewable sources in energy production increased, but fossil fuels continued to dominate. In particular, the high use of natural gas and coal led to an increase in carbon emissions. Turkiye's policies to comply with the Paris Agreement and its investments in renewable energy are considered as important steps towards reducing carbon emissions. Increasing the use of renewable energy sources and expanding energy efficiency projects can contribute to reducing Turkiye's carbon emissions. However, the dominance of fossil fuels in energy production and the continuation of the industrialization process will prevent the complete reduction of carbon emissions [21], [22]. In this regard, it is of great importance for Turkiye's energy policies to shift towards renewable energy sources and adopt environmental sustainability policies in line with sustainable development goals.

## LITERATURE REVIEW

Studies investigating environmental degradation and its causes focus on factors such as urbanization and industrialization, as well as their links to variables such as energy consumption, globalization, military expenditures, financial development and economic growth. The environmental impacts of these factors vary according to the characteristics of the country, region or sample analyzed. The literature provides a wide range of findings to understand the impacts of these variables on environmental degradation, thus contributing to the shaping of environmental policies at the regional and global level.

It is widely recognized that the environmental impacts of military activities, the increasing need for energy and the fossil fuels used in this process cause environmental damage [23], [24]. Military operations consume large amounts of fossil fuels, leading to increased carbon dioxide emissions and environmental damage. In addition, equipment such as military vehicles, helicopters, tanks and ships also consume large amounts of fossil and nuclear fuels. Researchers such as [25] and [7] emphasize that this situation contributes to increasing environmental degradation. In particular, [26] point out that the environmental impacts of military activities are not limited to the period in which they are carried out and lead to long-term ecological destruction. Encouraging renewable energy sources can offer a solution to these problems. Renewable energy sources such as wind, solar and hydropower contribute to improving air quality and reducing environmental pollution [27]. Increasing renewable energy consumption can significantly reduce per capita carbon emissions, especially in low-income countries. This demonstrates the potential of renewable energy to alleviate environmental pressures [28]. Furthermore, renewable energy projects can address issues related to clean water and sanitation systems, such as sustainable water supplies, water reuse, recycling and providing treatment facilities [29]. Renewable energy technologies are advantageous in terms of cost-effectiveness, safety and environmental sustainability and offer a more sustainable energy supply by reducing dependence on fossil fuels [30]. Therefore, the use of renewable energy to reduce the environmental impacts of military expenditures is of strategic importance for both national security and environmental sustainability.

Studies examining the relationship between energy consumption and environmental damage reveal the impacts of energy consumption on the environment. Kesbic and Bozduman [31] found that the increase in energy consumption in Turkiye increases $CO_2$ emissions and causes environmental pollution. Chontanawat [32] reported that there is a strong relationship between energy consumption and carbon emissions in ASEAN countries. Çetin and Yüksel [33] examined the relationship between energy consumption and carbon emissions in Turkiye within the framework of the Environmental Kuznets Curve (EKC) hypothesis and concluded that energy consumption increases carbon emissions. Al-Mulali and Sab [34] found a long-run relationship and bidirectional causality between electricity consumption and $CO_2$ emissions in Middle Eastern countries. Lee and Yoo [35], [36], in their separate studies for Mexico and Korea, observed bidirectional causality between energy consumption and $CO_2$ emissions. Doğan and Topallı [37] find bidirectional causality between energy consumption and $CO_2$ in Turkiye. Özkan and Erdemli [38], in their study on Turkiye and its border neighbors, showed that energy consumption has a positive relationship with $CO_2$ emissions in the long run.

The relationship between economic growth and environmental degradation has been analyzed in different aspects by various studies. The Environmental Kuznets Curve (EKC) hypothesis is a frequently used theory in these studies. Öztürk and Acaravcı [39] find that the EKC hypothesis is not valid for the period 1968-2005 in Turkiye. On the other hand, [40] find that the EKC hypothesis is valid for the period 1970-2010 in Algeria. Ahmad and Du [41], in their study for Iran, found that economic growth increases environmental pollution. Similarly, [42] showed that economic growth increases carbon emissions in Azerbaijan. Gökmenoğlu and Sadeghieh [43], in their long-run study on Turkiye, found that economic growth has a significant and negative impact on carbon emissions. The study by [44] in 12 MENA countries showed that economic growth causes environmental pollution. Chen et al. [45] found a unidirectional relationship between economic growth and carbon emissions in China and reported that economic growth continues to have negative impacts on the environment.

Studies investigating the impacts of financial development on environmental degradation reveal that financial development can have both positive and negative environmental consequences. Tamazian et al. [46] reports that high economic and financial development in BRIC economies improves environmental quality. Similarly, [47] finds that financial development in India reduces carbon emissions. Pata [48], in his analysis for Turkiye, showed that financial development increases carbon emissions along with energy consumption, urbanization and industrialization. Özkan and Erdemli [38] find a positive long-run relationship between energy consumption and environmental pollution in Turkiye and neighboring countries.

The environmental impacts of the globalization process vary between developed and developing economies. Shahbaz et al. [49] reported that globalization has negative impacts on environmental quality in developed economies. Apaydın [50] showed that globalization increased the ecological footprint of globalization in Turkiye but decreased the ecological footprint of exports. Wang et al. [51] reported that economic globalization increased environmental degradation in G7 countries, while the increase in the agricultural sector reduced carbon emissions. Karasoy [52] found that globalization, industrialization and urbanization have negative impacts on ecological footprint in Turkiye. Finally, [53] found that globalization reduces environmental damage in Turkiye, but urbanization increases environmental damage.

The rapid increase in urbanization leads to increased environmental damage, especially in developing countries. The impacts of urbanization on energy consumption, carbon emissions and ecological footprint are among the issues that have been researched. The environmental impacts of urbanization are intensively studied especially in developing countries. Yıldız and Göktürk [54] reported that urbanization creates a positive relationship between energy consumption and carbon dioxide emissions in Turkiye. Phong [55] showed that urbanization increases carbon dioxide emissions in ASEAN-5 countries and financial development supports this impact. Karasoy [52] found that urbanization, industrialization and energy consumption increase the ecological footprint in the short run in Turkiye. Apaydın [50] stated that the interactions between globalization and urbanization increase the ecological footprint, and this process triggers environmental degradation across Turkiye.

The literature on the environmental impacts of military expenditures generally analyzes the impacts of military expenditures on carbon emissions and environmental degradation. Ahmed et al. [56] examined the relationship between defense expenditures, energy consumption, carbon dioxide emissions and economic growth in Myanmar for the period 1975-2014. The results of the study show that defense expenditures have a negative impact on economic growth, while energy consumption supports growth. Ahmed et al. [57], in their study covering 22 OECD countries, found that defense expenditures increase carbon emissions. Similarly, [58] found a positive and significant relationship between military expenditures and $CO_2$ emissions in their study covering 120 countries. In a study conducted by [59] for 12 countries with the highest military expenditures, it was found that military expenditures have a positive impact on green growth in developed countries and a negative impact in less developed countries. A study by [25] on G7 countries finds a unidirectional causality between militarization and $CO_2$ emissions. In another study, [60] pointed out that militarization increases $CO_2$ emissions in the US. Erdoğan et al. [61], in their study for Greece, France, Italy and Spain, stated that military expenditures have serious negative impacts on the environment.

Kwakwa [62] finds that military expenditures, along with industrialization and public expenditures, increase $CO_2$ emissions in Ghana. Similarly, [63] revealed the positive impacts of defense spending on $CO_2$, $NO_2$ and methane emissions. Qayyum et al. [64] find that military expenditures and armed conflicts in South Asia led to ecological degradation. Zandi et al. [65] find that military expenditures increase $CO_2$ emissions in Far East Asian countries. On the other hand, there are few studies suggesting that military expenditures can improve environmental quality. For example, [66] argue that military R&D activities can reduce environmental degradation through technology innovation, while [67] argue that militarization in India can reduce carbon emissions.

The findings of studies conducted in Turkiye are similar to the general literature. Kurt and Kılıç [68] find that there is a long-run and bidirectional causality relationship between military expenditures and environmental damage. Gökmenoğlu et al. [6] found that military expenditures increase environmental degradation. Erdoğan [69], meanwhile, found no significant relationship between defense expenditures and carbon emissions in the long run, but showed that defense expenditures affect carbon emissions in the short run.

Contrary to the limited number of studies suggesting that military expenditures can improve environmental quality, it is generally found to cause environmental degradation and increase carbon emissions. However, how renewable energy sources can mitigate these negative impacts has not been examined in the literature. In this respect, this study investigates whether renewable energy sources can mitigate the environmental impacts of military expenditures in Turkiye.

## DATA AND METHODOLOGY

### Model Specification and Data

The empirical investigation of this study examines the potential role of renewable energy sources in reducing the environmental impacts of military expenditures in Turkiye. The study utilizes annual data for the period 1990-2021. This period was chosen based on the availability and suitability of the data for analysis. In our study, we used $CO_2$ emissions as the dependent variable and military expenditure, renewables energy, GDP per capita, population and interaction term as independent variables.

$$CO2_{(M-D)} = \alpha_0 + \beta_1 MILEX_t + \beta_2 RENEW_t + \beta_3 GDP_t + \beta_4 POP_t + \beta_5 INT_t + \mu_t \qquad (1)$$

The model of our study can be expressed in the following functional form:

In Equation 1 $\alpha_0$ is the constant term of the model. The coefficients $\beta_1$ to $\beta_2$ are slope coefficients that measure the impact of each independent variable on carbon dioxide emissions. $\mu_t$ is the error term of the model with zero mean and constant variance, where t denotes the time period.

Interaction term is a method used in statistical modeling to examine the combined impact of two or more independent variables on the dependent variable [70], [71]. Especially in multivariate regression analyses, it is used to indicate the interaction of independent variables with each other on the dependent variable beyond their direct impacts. In this study, the interaction term is used to analyze the potential role of renewable energy sources in reducing the environmental impacts of military expenditures in Turkiye. Thus, beyond the independent impacts of military expenditures and renewable energy use on carbon emissions, it measures the combined impact of these two factors. The indicators used in our study and their descriptive statistics are shown in Table 1.

**Table 1.** Descriptive statistics (1990-2021).

| Variable | Notation | Description | Mean | Median | Min. | Max. | Std. Dev. | Source |
|---|---|---|---|---|---|---|---|---|
| $CO_2$ emissions | $CO_2$ | metric tons per capita | 0.562 | 0.554 | 0.409 | 0.716 | 0.097 | WB |
| Military expenditure | MILEX | constant US$ | 3.968 | 3.951 | 3.802 | 4.210 | 0.102 | SIPRI |
| Renewables energy | RENEW | % of primary energy | 1.106 | 1.111 | 0.915 | 1.287 | 0.091 | EIA |
| GDP per capita | GDP | constant US$ | 11.732 | 11.735 | 11.461 | 12.054 | 0.184 | WB |
| Population | POP | total | 0.147 | 0.143 | -0.099 | 0.277 | 0.082 | WB |
| Interaction term | INT | Military*Renewable | 4.794 | 4.776 | 4.210 | 5.344 | 0.377 | AUT |

**Note:** (1) WB, SIPRI, EIA, AUT indicate World Bank-World Development Indicators, Stockholm International Peace Research Institute, U.S. Energy Information Administration, and Author's Calculated Data respectively. (2) All variables were logarithmized.

Table 1 shows that the average $CO_2$ emissions per capita is 0.562 metric tons. The median value is also close to the mean at 0.554, indicating a symmetrical distribution of the data. The standard deviation of 0.097 suggests that the data are closely clustered around the mean and hence show low variability. Military expenditures are calculated in constant US dollars, with a mean of \$3,968. The median is \$3,951, again indicating a symmetrical distribution of the data. The standard deviation is 0.102, which is a low value and indicates stability in military expenditures over the years.

The average share of renewable energy in primary energy is 1.106%. The median value of 1.111% is close to the average value and indicates a homogeneous distribution of renewable energy use in general. Renewable energy utilization rates range from a minimum of 0.915% to a maximum of 1.287%. These values show the diversity in renewable energy use across periods. The standard deviation of 0.091 indicates that these rates are generally similar and do not show large fluctuations. The average GDP per capita is 11,732 dollars. The median value is very close to the mean value at \$11,735. This suggests that GDP values generally show a central trend. The standard deviation is 0.184, indicating a moderate level of variability, but there is still limited variability. Finally, the population size has a mean of 0.147. The median value of 0.143 is close to the mean value, indicating a symmetrical distribution of the data. The standard deviation of the population variable is 0.082, indicating low variability and a generally stable population size.

**Unit Root Analysis**

In time series analysis, it is important to determine whether the series are stationary in order to make robust estimates. Especially economic time series often exhibit non-stationary characteristics [72]. This leads to spurious regressions, causing biased and inconsistent estimates [73]. Therefore, Augmented Dickey-Fuller (ADF), Phillips-Perron (PP) and Kwiatkowski-Phillips-Schmidt-Shin (KPSS) unit root tests are used in this study. The application, rationale and explanation of the results of each of these tests are analyzed in the rest of the paper.

Augmented Dickey-Fuller (ADF) Unit Root Test: The ADF test developed by [74] is used to determine whether a time series is stationary. The ADF test accepts the null hypothesis that there is a unit root in the series. The alternative hypothesis is that the series is stationary. The test statistic measures the relationship between the previous and current values of the dependent variable in regression analysis. During the application of the test, the lag length is determined according to the model selection criteria (AIC, SIC, etc.). The general structure of the ADF test is shown in Equation 2 [74]:

$$\Delta y_t = \alpha + \beta_t + \gamma y_{t-1} + \sum_{i=1}^{p} \varphi i \Delta y_{t-1} + \varepsilon_t \quad (2)$$

In Equation 2 $y_t$ is the series itself, $\Delta y_t$ is the first difference of the series, $y_{t-1}$ is the previous value of the series, $\alpha$ is the constant term, $\beta_t$ is the trend term and $\varepsilon_t$ is the error term. The null hypothesis of the ADF test is that the series has a unit root.

Phillips-Perron (PP) Unit Root Test: In contrast to the ADF test, the PP test developed by [75] makes additional corrections to model autoregressive and moving average structures in the series. The PP test uses nonparametric methods to correct for autocorrelation and heteroskedasticity in the series. This makes the test statistics more robust. The advantage of the PP test is that it eliminates the need to decide the lag structure of the model and is thus easier to use [75]. The formulation of the PP unit root test is shown in Equation 3:

$$y_t = \rho y_{t-1} + \varepsilon t \quad (3)$$

In Equation 3, the coefficient ρ is the correlation of the series with its previous values. εt is the error term and takes into account autocorrelation and heteroskedasticity. Both the ADF and PP tests suggest that the null hypothesis (the existence of a unit root in the series) should be rejected when the calculated test statistics are greater than the critical values.

Kwiatkowski-Phillips-Schmidt-Shin (KPSS) Test: The main difference of the test developed by [76] is that the null hypothesis assumes that the time series is stationary. The KPSS test tests stationarity by checking whether the series has a permanent trend. The test has the ability to test the level and trend stationarity of the series separately. In the application of the KPSS test, one of the trend or level stationarity options should be preferred in accordance with the structural characteristics of the series. The interpretation of the test results is that the series contains a unit root if the null hypothesis is rejected [76]. The regression model of the test is shown in Equation 4:

$$y_t = \mu + \tau_t + \varepsilon t \quad (4)$$

In Equation 4, μ is the constant term, $\tau_t$ is the deterministic trend and $\varepsilon_t$ is the error term. The variance of the cumulative sum of the error terms is tested.

**ARDL Cointegration Test**

The Autoregressive Distributed Lag (ARDL) bounds testing method developed by [77] offers significant advantages over classical cointegration tests such as [72], [78], [79]. The ARDL approach allows the variables under study to be stationary at level or first differences, which means that the model can include both I(0) and I(1) series.

The ARDL model is more suitable for small sample sizes. It also allows the evaluation of short-run and long-run dynamics in the same model. This feature is ideal for investigating the existence of long-run relationships. The ARDL cointegration test is constructed to include past values of the dependent variable and both past and current values of the explanatory variables. The ARDL model is as in Equation 5 [77]. This equation represents the long-run cointegration relationship between short-run dynamics and the error correction term.

$$Y_t = a_0 + \sum_{i=1}^{p} a_i Y_{t-1} + \sum_{j=0}^{q} \beta_{j1} X_{1,t-j} + \cdots + \sum_{j=0}^{q} \beta_{jk} X_{k,t-j} + \epsilon_t \qquad (5)$$

The optimal lag length in the model is determined using selection criteria such as Akaike Information Criterion (AIC) and Schwarz Information Criterion (SIC). The bounds determined by [77] are compared with the F-statistic to test for cointegration. If the F-statistic exceeds the upper critical value, cointegration between the variables is accepted. If the F-statistic is below the lower critical value, it is concluded that there is no cointegration. The robustness of the model is assessed by normal distribution, autocorrelation, variance and functional form tests. The stability of the long-run parameters is determined by the CUSUM and CUSUMQ tests developed by [80].

After cointegration is detected with the ARDL method, co-efficients are estimated. In this process, long-run coefficients are calculated first. After the long-run coefficients, short-run coefficients are calculated. At this stage, a change is made in the configuration of the model and the error correction term is included in the model. This term is derived from one lagged value of the residuals of the model in which the long-run relationship was identified. The error correction term indicates how much of the short-run imbalance will be corrected in the long run. The general expectation is that this coefficient should be negative and significant. Because this indicates the capacity of the model to rapidly stabilize towards the long-run stability [81], [82].

## EMPIRICAL FINDINGS

In this stage of our study, unit root tests, which is an important stage of the econometric modeling process, were applied. Unit root tests were applied for the six variables used in the study and their stationarity was analyzed. These variables were evaluated within the framework of both constant term and constant and trend term models. The results of these tests are shown in Table 2.

Table 2 shows that most of the variables are non-stationary in the ADF and PP tests, i.e. they contain unit roots. This implies that the original series do not have a constant mean and variance over time. However, when the first differences of the variables are taken, the obtained t-statistics are usual-

**Table 2.** Unit root test results

|  |  | ADF unit root test | | PP unit root test | | KPSS unit root test | |
| --- | --- | --- | --- | --- | --- | --- | --- |
| Variable |  | t-statistic (level) | t-statistic (first difference) | t-statistic (level) | t-statistic (first difference) | t-statistic (level) | t-statistic (first difference) |
| $CO_2$ | Constant | -0.866 | -6.007*** | -0.821 | -8.439*** | 0.734 | 0.206*** |
| MILEX |  | -0.630 | -3.957*** | -0.893 | -3.939*** | 0.534 | 0.119*** |
| RENEW |  | -2.088 | -6.089*** | -2.088 | -6.584*** | 0.174*** | 0.176*** |
| GDP |  | 0.514 | -5.509*** | 1.716 | -6.344*** | 0.745 | 0.333*** |
| POP |  | -2.050 | -6.827*** | -1.727 | -10.519*** | 0.612* | 0.500** |
| INT |  | -1.480 | -6.117*** | -1.388 | -6.288*** | 0.493** | 0.227*** |
| $CO_2$ | Constant and Trend | -3.163 | -5.929*** | -3.128 | -7.855*** | 0.065 | 0.175** |
| MILEX |  | -1.263 | -3.866** | -1.638 | -3.845** | 0.115* | 0.112*** |
| RENEW |  | -2.122 | -3.411** | -2.122 | -6.725*** | 0.160 | 0.131*** |
| GDP |  | -2.582 | -5.479*** | -2.594 | -7.065*** | 0.155* | 0.206* |
| POP |  | -3.334* | -6.733*** | -3.268* | -9.490*** | 0.127** | 0.484** |
| INT |  | -1.498 | -5.860*** | -1.407 | -6.204*** | 0.153* | 0.083*** |

**Note:** The superscripts ***, **, and * denote the significance at a 1%, 5%, and 10% level, respectively.

ly statistically significant. This indicates that taking the first differences makes the series stationary. Thus, the series are suitable for econometric modeling. The KPSS test works under the assumption that the series are stationary. The results of this test also confirm that most of the series are non-stationary at the first level but become stationary when first differences are taken. The results of the KPSS test support the findings of the ADF and PP tests. In the next stage of this study, ARDL bounds test is used to analyze the cointegration relationships between the series. The results of the ARDL bounds test are presented in Table 3.

**Table 3.** The results of ARDL cointegration test

| Model | Optimal lag length | F-statistics | Critical values %5 | | Critical values %1 | |
|---|---|---|---|---|---|---|
| | | | I(0) | I(I) | I(0) | I(I) |
| Model: F($CO_2$| MILEX, RENEW, GDP, POP, INT) | (1, 1, 0, 0, 1) | 6.694*** | 2.56 | 3.49 | 3.29 | 4.37 |

**Note:** The superscripts ***, **, and * denote the significance at a 1%, 5%, and 10% level, respectively.

Table 3 reveals that the structural formulation of the model and the optimal lag lengths are determined with the help of information criteria (AIC and SIC). The optimal lag length of the model is (1, 1, 1, 0, 0, 0, 1) and this structure is designed to capture the lagged impacts in the time series of the models in the best way. The F-statistic of the model is calculated as 6.694. The F-statistic exceeds the critical values at 5% and 1% significance levels. The results of the analysis indicate that there is a long-run and statistically significant relationship between the variables. This paved the way for estimating the long-run coefficients in the later stages of the study. Various specification tests were conducted to assess the econometric robustness of the model. These tests include the LM test for serial correlation, histogram-normality test, heteroskedasticity test and Ramsey RESET test. In addition, CUSUM and CUSUM of Squares tests were applied to detect structural breaks. The results of these tests reveal that there is no specification error in the model and the short and long-run estimations of the models are robust. The results of specification tests and long-run coefficient estimates are presented in Table 4, while the results of CUSUM and CUSUMSQ tests are presented in Figure 3.

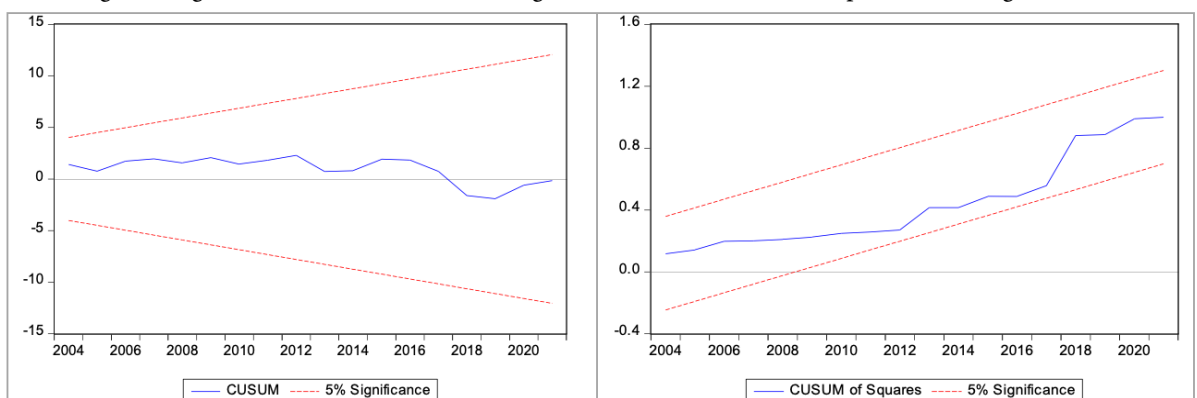

**Figure 3.** Plots of CUSUM and CUSUMSQ statistics.

**Table 4.** Short-run and long-run results

| Dependent variable: $CO_2$(M-D) | | |
|---|---|---|
| Regressors | Short-run coefficients | Long-run coefficients |
| MILEX | 0.260*** | 0.196*** |
| RENEW | -0.119** | -0.120*** |
| GDP | 0.162** | 0.193*** |
| POP | 0.047 | 0.095*** |
| INT | -0.130*** | -0.140*** |
| ECT(-1) | -0.445*** | |
| C | | -1.947** |
| Diagnostic tests | | P value |
| $x^2$ (Serial correlation) | | 0.58 |
| $x^2$ (Heteroskedasticity) | | 0.44 |
| $x^2$ (Normality) | | 0.55 |
| $x^2$ (Functional form) | | 0.73 |
| CUSUM | | Stable |
| CUSUMSQ | | Stable |

**Note:** The superscripts ***, **, and * denote the significance at a 1%, 5%, and 10% level, respectively.

Table 4 indicates that the error correction term (ECT) is negative and statistically significant. This result means that there is a long-run equilibrium relationship between the model variables and that short-run deviations will be eliminated in the long run. The value of the ECT coefficient (-0.445) indicates that approximately 45% of the short-run deviations will be corrected in the next period.

The results demonstrate that military expenditures have a positive impact (with coefficient of 0.260) on $CO_2$ emissions in the short run. Carbon emissions increase especially due to the use of fossil fuels by military vehicles and facilities. The impact of military expenditures on $CO_2$ emissions continues in the long run (with coefficient of 0.196). Military expenditures increase the defense capability of the country. However, the environmental impacts of this situation should also be taken into consideration. Therefore, it is very important to take steps towards the use of sustainable energy resources in military planning. On the other hand, the fact that the use of renewable energy reduces $CO_2$ emissions both in the short (with coefficient of -0.119) and long run (with coefficient of -0.120) proves that these energy sources are environmentally friendly. In addition to reducing carbon footprint, renewable energy offers a sustainable alternative to fossil fuels. From an economic perspective, investments in renewable energy technologies can provide economic benefits by reducing energy costs in the long run and have the potential to create jobs. On the other hand, the positive impact of GDP on $CO_2$ emissions (coefficients of 0.162 and 0.193 in the short and long run, respectively) suggests that economic growth is generally associated with increased energy consumption and industrial activity. This may increase the environmental pressure of the growth process. Adopting greener growth strategies, increasing energy efficiency and promoting sustainable industrial practices are needed to reduce the environmental impacts of economic growth. While the impact of population growth on $CO_2$ emissions is not statistically significant in the short run, it is positive in the long run (with coefficient of 0.095). Population growth leads to an increase in energy demand. As a result, environmental pressure increases and more $CO_2$ emissions are released. Therefore, economic and social policies should be sustainable in terms of population, urbanization and infrastructure. The results obtained in the study are consistent with studies such as [25], [57], [58], [60], [61].

Moreover, the interaction term is used to measure the combined impact of military expenditures and renewable energy use on $CO_2$ emissions in Turkiye. The results show that both the short-run (-0.130) and long-run (-0.140) coefficients of the interaction term are negative. These negative coefficients suggest that the use of renewable energy has the potential to reduce the negative impacts of military expenditures on the environment. It also indicates that this interaction offers an important opportunity for environmental sustainability. Military expenditures are generally energy-intensive activities and this energy use is usually based on fossil fuels. The combustion of fossil fuels releases gases such as $CO_2$ into the atmosphere. The findings of this study show that increasing renewable energy sources can reduce the environmental impact of military activities. The results imply a greater use of renewable energy sources, especially during military operations or in military installations. For example, solar panels could be installed on military bases or alternative energy sources could be developed for military vehicles. In the long run, the persistent negative impact of the interaction term highlights that such policies and technological integrations can increase their impact over time and underscores the importance of taking concrete steps towards sustainability goals in the military sector. Moreover, the positive consequences of this interaction will create a strategic synergy between national security and environmental policies. Considering environmental factors in military planning can strengthen Turkiye's environmentally friendly image in the international arena. Thereby, Turkiye's reputation for environmental sustainability could be enhanced and investments in green technologies could increase.

## CONCLUSION

It is a widely accepted fact that military expenditures increase $CO_2$ emissions and environmental pressure. On the other hand, renewable energy sources are seen as sustainable and environmentally friendly alternatives and have the potential to reduce environmental pressures. Therefore, this study investigates whether renewable energy sources can reduce the environmental impact of military expenditures in Turkiye by using the ARDL method. To measure this impact, an interaction term is included in the ARDL model.

The findings of the study show that military expenditures in Turkiye have a significant impact on $CO_2$ emissions; military expenditures increase $CO_2$ emissions in both the short and long run. This is attributed to the high energy consumption of military activities and the fact that most of this energy is provided by fossil fuels. However, the use of renewable energy sources significantly reduces this negative impact. In both the short and long term, the use of renewable energy has been found to reduce $CO_2$ emissions. This suggests that renewable energy should be more widely adopted in the military sector. In addition, economic growth and population growth are also found to increase $CO_2$ emissions. These results reveal that Turkiye needs to make strategic moves to ensure environmental sustainability in its development process. In particular, it is important to implement policies that will balance the environmental impacts of economic growth and population growth. Finally, interaction term analysis reveals that the use of renewable energy has the potential to reduce $CO_2$ emissions from military expenditures. This potential will play a supportive role for the renewable energy sector while reducing the environmental impacts of military expenditures.

Based on the results, the use of renewable energy sources can alleviate the pressure of military expenditures on the environment. Therefore, a series of policy recommendations have been prepared to support the reduction of $CO_2$ emissions in Turkiye: (1) Turkiye should develop policies to promote the use of renewable energy in its military and civilian infrastructure. The installation of solar panels and wind turbines on military installations should be supported and the use of electric or hybrid models for military vehicles should be encouraged. These steps would reduce the environmental impact of military expenditures and stimulate the renewable energy sector. (2) As part of efforts to reduce the environmental impact of military expenditures, mandatory environmental impact assessments should be conducted for military projects. These assessments will identify the potential impacts of projects on the environment and enable the development of strategies to minimize negative impacts. (3) Turkiye should align its economic growth strategies with the principles of environmental sustainability and encourage the use of environmentally friendly technologies in industrial production. In addition, the adoption of new energy efficient technologies and the promotion of environmentally friendly innovations should be a fundamental part of economic policies. (4) Urban planning and infrastructure projects should be designed in line with sustainable development objectives to manage the environmental impacts of population growth. Environmentally friendly practices such as energy-efficient buildings and waste management systems should be encouraged. (5) Increase cooperation between the military and civilian sectors on the use of renewable energy. These collaborations will accelerate the adoption of renewable energy technologies in both sectors and have the potential to have a widespread environmental impact.

Although this study provides important findings, it has some limitations. Future studies can address these limitations. Firstly, in this study, environmental pressure is measured by $CO_2$ emissions per capita. In future studies, other variables representing environmental pressure can be added to the model or an environmental pressure index can be constructed with these variables. Secondly, future studies can expand the sample to include both developed and developing countries. In this way, it can be compared whether the results differ in countries with different income levels. Thirdly, by using different econometric models, graphs reflecting the effectiveness of the use of renewable energy sources in reducing environmental impacts can be created. Finally, larger data sets and different analysis techniques can be used to test the consistency of the findings of this study.

## DATA AVAILABILITY STATEMENT

The author declared that the data that supports the findings of this study are available within the article. Raw data that support the finding of this study are available from the corresponding author, upon reasonable request.

## CONFLICT OF INTEREST

The author declared that there are no potential conflicts of interest with respect to the research, authorship, and publication of this article.

## USE OF AI FOR WRITING ASSISTANCE

The author declared that AI was not used in writing or as a writing assistant in the manuscript.

## ETHICS

The author declared that there are no ethical issues related to the publication of this article.

## FUNDING STATEMENT

The author declared that the article has not received financial support.